\documentclass[prl,amsmath,amssymb,superscriptaddress,twocolumn]{revtex4-2}

\usepackage{amsmath}
\usepackage{amssymb}
\usepackage{amsthm}
\usepackage{makecell}
\usepackage{amsbsy}
\usepackage{amstext}
\usepackage{graphicx}
\usepackage[dvipsnames]{xcolor}
\usepackage{float}

\usepackage{wasysym}

\usepackage{mathtools}

\makeatletter
\usepackage{amsfonts}
\usepackage{dcolumn}
\usepackage{bbold,bm}
\usepackage{hyperref}
\hypersetup{
    colorlinks=true,
    linkcolor=blue,
    filecolor=magenta,      
    urlcolor=cyan,
    pdfpagemode=FullScreen,
    }

\newcommand\vex[1]{\mathbf{#1}}
\newcommand\gvex[1]{\boldsymbol{#1}}

\def\ket#1{\mathinner{|{#1}\rangle}}

\makeatother
\begin{document}
  \title{Tunable Non-Equilibrium Magic and Minimum Twist Angles in AA-Stacked Twisted Multilayer Graphene}

\author{Yantao Li}
\affiliation{Department of Physics and Astronomy, The University of Alabama, Tuscaloosa, AL 35487, USA}
\author{Wang-Kong Tse}
\affiliation{Department of Physics and Astronomy, The University of Alabama, Tuscaloosa, AL 35487, USA}

\begin{abstract}
We report the discovery of a series of non-equilibrium magic angles at which isolated topological flat quasienergy bands form in AA-stacked twisted multilayer graphene under circularly polarized light. These non-equilibrium magic angles can be traced back to specific static twist angles where the bandwidth reaches a minimum \textit{without} the formation of isolated flat bands. We refer to these as minimum twist angles, in contrast to the magic angles observed in twisted bilayer graphene. We show that an applied displacement field can further flatten the optically induced topological flat bands accompanied by larger non-equilibrium magic angles. The discovery of these electrically tunable topological flat quasienergy bands is expected to open up a new avenue of exploring exotic Floquet-driven phenomena in AA-stacked twisted multilayer graphene. 
\end{abstract}

\date{\today}

{
\let\clearpage\relax
\maketitle
}

\emph{Introduction}.---Moir\'e graphene superlattice has attracted intense interest since the experimental discovery of correlated insulating states and superconductivity in twisted bilayer graphene (TBG)~\cite{cao2018correlated,cao2018unconventional}. Since then, moir\'e systems have been experimentally realized in materials such as twisted multilayer graphene (TMG) structures~\cite{yankowitz2019tuning,Sharpe_2019,Fleischmann_2020,Jones_2020,cao2021large,Xu_2021,Choi_2021,Park2022,uri2023superconductivity,su2024} as well as other platforms~\cite{Naik_2018,Tang2020,Wang2020,Zhang_2020,Shabani_2021,Xu2022,Xiong2022,Meng2023}. Theoretically, their unique band structures and interaction effects have been predicted to give rise to a host of interesting states, including 
Mott insulators, quantized anomalous Hall insulators, and $d$-wave unconventional superconductors~\cite{Xu_2018,Po_2018,Fengcheng_2018,Kang_2018,Kang_2019,Seo2019,Koshino_2019,Hejazi_2019,Chebrolu_2019,Cea_2019,Liu_2019,Liang_2020,Tran_2020,Tritsaris_2020,Lake2021,Qin2021,Eaton2022,Foo2024,Zhou2022,Valagiannopoulos2022,Adhikari2023,Song2022,lau2023topological,Li2024}.

In TBG, two single graphene sheets are twisted relative to each other at a certain angle. At the so-called flat-band magic angles, 
the dispersion of the Dirac cones in each layer becomes flat, forming isolated flat bands 
~\cite{Bistritzer_2011,dosSantos_2007,Shallcross_2008,Shallcross_2010,Mele_2010,Mele_2011,Carr_2017}. This can be theoretically described by the single-particle Bistritzer and MacDonald's (BM) model. 
The origin of the magic angles can be traced back to the limit when the interlayer tunneling of the electrons in the AA-region of TBG is suppressed, known as the chiral limit ~\cite{Tarnopolsky_2019}. In alternating TMG, similar magic angles occur with additional Dirac cones superposed on the flat bands at the $K$ or $K^{\prime}$ points~\cite{Khalaf_2019}.

Such isolated flat bands provide opportunities to experimentally study a myriad of fascinating phenomena such as anomalous Hall effects~\cite{Sharpe_2019,Serlin2020,Tseng2022} and Chern insulators~\cite{Nuckolls2020,Stepanov2021} in TBG. In particular, fractional Chern insulators have been experimentally observed~\cite{Xie2021} as the flat bands can effectively mimic Landau levels.

Notably, in AA-stacked twisted multilayer graphene, such isolated flat-band magic angle phenomena do not occur, limiting the motivation to explore exotic physics in these systems. However, another type of band reconstruction exists, known as the Dirac magic phenomenon, where multiple highly anisotropic Dirac cones coexist at the $K$ point~\cite{Li2022}. These special twist angles are called Dirac magic angles. Unlike the flat-band magic angles that originate from the chiral limit, Dirac magic angles have a geometric origin~\cite{Li2022}.

Under circularly polarized laser illumination, the quasienergy structures of periodically driven TBG, alternating TMG or other types of TMGs display two isolated topological flat bands with a gap opened by time-reversal symmetry breaking~\cite{Li_2020,Katz_2020,Vogl2020,Martin2020,Lu2021,Vogl2021,Assi2021,Hu2023,Huang2023,dubey2024}. It should be stressed that such isolated topologically non-trivial flat bands originate from the equilibrium flat bands that already exist at the magic angles before illumination, and so the effects of light are simply to further flatten the bands and to open up a gap. 

In this paper, we uncover a series of twist angles at which isolated topologically non-trivial flat bands emerge under specific driving amplitudes and frequencies in AA-stacked TMG. These non-equilibrium magic angles (NEMA) originate from particular twist angles as the system approaches equilibrium. In contrast to flat-band magic angles and Dirac magic angles, these twist angles correspond to the minimum bandwidth values that do \textit{not} host any isolated flat bands, which we refer to as minimum twist angles (MTA). To demonstrate NEMA and MTA, we derive an analytic formula within the high-frequency approximation and numerically compute the Floquet band structure. Additionally, we apply a displacement electric field and show that both NEMA and MTA can be finely tuned by the electric field.

\begin{figure}[t]
   \centering
   \includegraphics[width=3.4in]{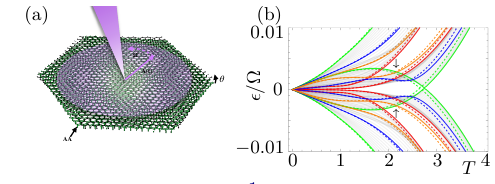} 
   \caption{(a) Schematic illustration of irradiated AA/S-stacked twisted multilayer graphene. (b) Quasienergy band structure displaying two central isolated topologically non-trivial flat bands (indicated by black arrows). The driving amplitude is $A_0=0.45$ ($\hbar/ea$) with a twist angle of $1.22^\circ$. The driving frequency $\Omega\equiv2\pi/T$ and the unit of $\Omega$ is $\hbar v_F / a = 2.4$ eV. Dashed and solid lines represent analytic and exact numerical results, respectively, while colored lines denote symmetry points.}
   \label{fig1}
\end{figure}

\emph{Set up and model of AA-stacked TMG}.---The set up of the irradiated TMG is shown in Fig.~\ref{fig1}(a). The TMG (denoted as AA/D)  consists of an AA-stacked bilayer graphene (AA) twisted relative to another graphene system (D). Following the Bistritzer-MacDonald model~\cite{Bistritzer_2011}, the system Hamiltonian can then be written as ~\cite{Li2022}
\begin{equation}\label{eq:HMG1}
H_{\text{AA/D}}=
\begin{bmatrix}
 h_{\text{AA},\theta/2} & T \\
T^\dagger & h_{\text{D},-\theta/2} \\
\end{bmatrix},
\end{equation}
where
\begin{equation}\label{eq:HMG2}
h_{\text{AA},\theta/2}=
\begin{bmatrix}
 h_{\theta/2}+V\mathbb{1}_{2\times2} & T_{\text{AA}} \\
T_{\text{AA}}^\dagger & h_{\theta/2} \\
\end{bmatrix},
\end{equation}
and $h_{\theta/2}=-v_Fi\hbar \gvex\nabla \cdot\gvex\sigma_{\theta/2}$, with Fermi velocity $v_F$ and rotated Pauli matrices $\gvex\sigma_{\theta/2} \equiv e^{-i\theta\sigma_z/4} (\sigma_x,\sigma_y) e^{i\theta\sigma_z/4}$. $V$ is the potential from a vertically applied displacement field. In the main text of this paper, we take D to be a single-layer graphene (S), and consider the TMG system AA/S. Additional results for AA/AB and AA/ABC TMGs in support of our findings are presented in the Supplemental Material~\footnote{See Supplemental Material for the NEMAs and MTAs for AA/AB and AA/ABC TMGs.}. Then, we have $h_{\text{D},-\theta/2}=h_{-\theta/2}-V\mathbb{1}_{2\times2}$. The interlayer tunneling matrix for AA-stacked bilayer graphene is~\cite{Lobato2011} 
\begin{equation}\label{eq:TAA}
T_{\text{AA}}=
\begin{bmatrix}
\gamma^1_{\text{AA}} & \gamma^4_{\text{AA}}\gvex k \\
\gamma^{4}_{\text{AA}}\gvex k ^{\star} & \gamma^1_{\text{AA}} \\
\end{bmatrix},
\end{equation}
where $\gvex k=k_{x}+i k_{y}$ and the off-diagonal matrix elements which are $\gvex k$ dependent representing trigonal warping and particle-hole asymmetry. The interlayer tunneling matrix for TBG is $T = \sum_{n=1}^3 T_n e^{-i k_\theta \vex q_n \cdot \vex r}$, with
\begin{equation}
T_n = w_{\text{AA}} \sigma_0 + w_{\text{AB}} \vex q_n \cdot \gvex\sigma_{\pi/2},
\end{equation}
where the unit vectors $\vex q_1 = (0,-1)$ and $\vex q_{2,3} =  (\pm\sqrt3/2,1/2)$. $w_{\text{AA}}$ and $w_{\text{AB}}$ are the tunneling amplitudes between the AA- and AB-stacked regions of the TBG. We set the lattice relaxation parameter $u=w_{\text{AA}}/w_{\text{AB}}$~\cite{Carr2019}. Also, $k_\theta = 8\pi \sin(\theta/2)/3 a$ is the wave vector of the moir\'e pattern and $a$ is the Bravais lattice spacing of graphene. 

The values of the parameters introduced above are given as follows: $a= 2.4$~\AA, $\hbar v_F / a = 2.4$~eV, and $w_{\text{AB}} = 110$~meV. For AA-stacked bilayer graphene, $\gamma^1_{\text{AA}}=217$ meV and $\gamma^4_{\text{AA}}/a=-20$ meV. 

\begin{figure}[t]
   \centering
   \includegraphics[width=3.4in]{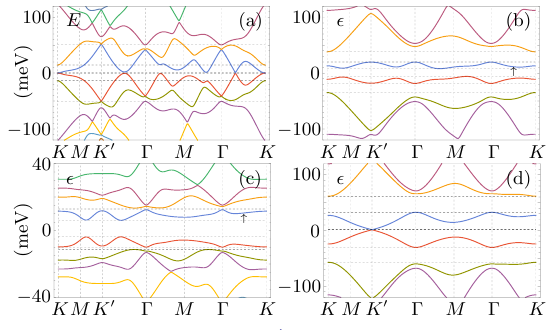} 
   \caption{Exact numerical band structures. (a) shows the static bands at $1.22^\circ$. (b) and (c) show the quasienergy band structure for the twisted angle $\theta^{f}_1=1.22^\circ$ and $\theta^{f}_2=0.68^\circ$ in the first and second range of non-equilibrium magic angles in which two central isolated topological flat bands form (pointed by up arrows). In (b), $T=2.18$ and $A_0=0.45$. In (c), $T=1.2$ and $A_0=0.45$. In (d), $T=2.77$, $A_0=0.45$, and $\theta^{f}_1=1.22^\circ$ in which the two central bands touch at $K^{\prime}$ point indicating a topological phase transition happens. All of the plots do not have a displacement electric field.}
   \label{fig2}
\end{figure}

\begin{figure*}[ht]
   \centering
   \includegraphics[width=7.2in]{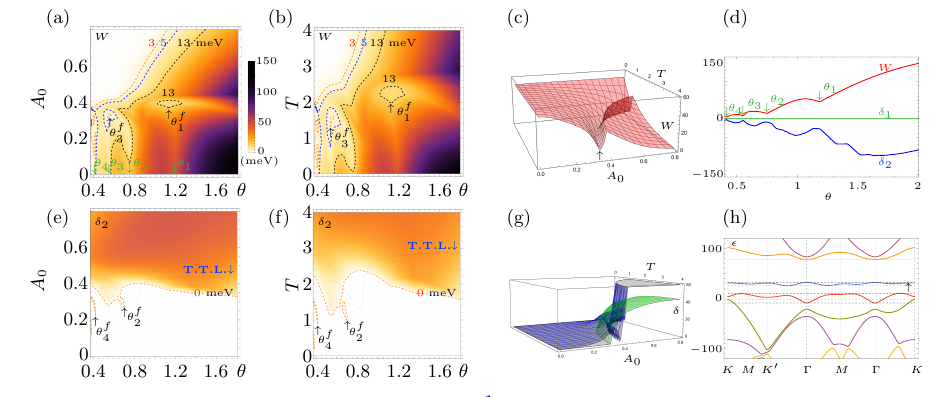} 
   \caption{Exact numerical solutions. (a), (b), and (c) show the quasienergy bandwidth ($W$) of the first positive central band. $\theta^{f}$s are the angles in the range of non-equilibrium magic angles (up arrows in (a) and (b)) and $\theta$s are twist angles (minimum twist angles) of the corresponding static origin (down arrows). In (a), $T=3$. In (b), $A_0=0.45$. In (c), the up arrow points to the positions of an isolated band at $1.22^\circ$. (e), (f), and (g) show the quasienergy band gaps ($\delta$) between the two central bands (blue) and between the first positive band and the second positive band (green), and corresponding (a), (b), and (c) separately. T.T.L. means topological transition line ($\delta_1=0$), and the topological trivial region is above the line. (d) shows the static bandwidth $W$ of the first positive band, band gap $\delta_1$ between two central bands, and band gap $\delta_2$ between the second positive and the first positive band (the second minus the first band). (h) shows the quasienergy band structure (solid lines) of $\theta^{f}_{1}=1.35^\circ$ in the first range of non-equilibrium magic angles with displacement electric field $\Delta V=400$ meV in which a topological flat band forms with bandwidth $W=6.8$ meV (pointed by an up arrow). In (h), $A_0=0.45$, $T=3.38$. All of the plots except (h) do not have a displacement electric field.}
   \label{fig3}
\end{figure*}

\emph{Floquet theory and analytical results}.---When a system is periodically driven by an external field, the Hamiltonian is time-periodic with a period $T\equiv2\pi/\Omega$, allowing the use of Floquet theory~\cite{Sam73a,Oka2009,Kitagawa2011,Lindner2011,Jiang2011,Li2014,Kundu2014,Else2016,Ke2020,Kuhlenkamp2020}. Due to discrete-time translation symmetry, the solution of the Schr\"odinger equation can be written as $\ket{\psi_{\bold{k},\alpha}(t)} = e^{-i\epsilon_{\bold{k},\alpha}t}\ket{\phi_{\bold{k},\alpha}(t)}$, where $\epsilon_{\bold{k},\alpha}$ is the quasienergy and $\ket{\phi_{\bold{k},\alpha}(t+T)}=\ket{\phi_{\bold{k},\alpha}(t)}$ is the time-periodic Floquet-mode wave function. They satisfy the Sch\"odinger-Floquet equation $[H_{\bold{k}}(t)-i\partial_t]\ket{\phi_{\bold{k},\alpha}(t)}=\epsilon_{\bold{k},\alpha}\ket{\phi_{\bold{k},\alpha}(t)}$~\cite{Sam73a}. $\epsilon_{\bold{k},\alpha}$ can be folded back from other Floquet Brillouin zones and then restricted to the first Floquet Brillouin zone $(-\Omega/2,\Omega/2]$. The quasienergies and the Floquet-mode wave functions can be exactly obtained from solving
\begin{equation}\label{timeE}
U(\bold{k},T,0)\ket{\phi_{\bold{k},\alpha}\left(0\right)}=\exp\left(  -i\epsilon_{\bold{k},\alpha}T\right) \ket{\phi_{\bold{k},\alpha}\left(0\right)},
\end{equation}
where $U(\bold{k},t,t') = \mathcal{T} \exp\left(-i\int_{t'}^t d\tau H(\mathbf{k},\tau) \right)$ is the time-evolution operator. In the high-frequency limit, we can also obtain approximate quasienergies and the Floquet-mode wave functions. Expanding the time-periodic Hamiltonian and Floquet-mode  wave function in the Fourier space as $H(t)=\sum_{m}e^{im\Omega t}H^{(m)}$ and $|\phi_\alpha(t)\rangle=\sum_me^{im\Omega t}|\phi_\alpha^{(m)}\rangle$, the Schr\"odinger-Floquet equation takes the form  $H_{\mathrm{eff}}|\phi_\alpha^{(0)}\rangle = \epsilon_\alpha|\phi_\alpha^{(0)}\rangle$, where  $H_{\mathrm{eff}} = H^{(0)} + \left[H^{(-1)},H^{(1)}\right]/\Omega$ is the effective Floquet Hamiltonian up to first order in $1/\Omega$~\cite{Kitagawa2011}. 

The AA-stacked TMG considered in this work is irradiated with circularly polarized light described by the vector potential $\vex A(t)=A_0(\cos(\Omega t),\sin(\Omega t))$. Then, minimal coupling in the BM model Hamiltonian Eq.~\eqref{eq:HMG1} gives $h_{\pm\theta/2}(t) = v_F[-i\hbar \gvex\nabla - e \vex A(t)]\cdot\gvex\sigma_{\pm\theta/2}$. We then find that $H_{\mathrm{eff}} - H^{(0)}$ equals
\begin{equation}\label{Eff1}
\begin{aligned}
\frac{(ev_F A_0)^{2}}{\Omega}\begin{bmatrix}
1+(\gamma^4_{\text{AA}})^{2}& 2\gamma^4_{\text{AA}} & 0\\
2\gamma^4_{\text{AA}}&1+(\gamma^4_{\text{AA}})^{2}&0\\
0 & 0 &1 \\
\end{bmatrix}
\mathbb{1}_{2\times 2}\otimes\hat{\sigma}_z.
\end{aligned}
\end{equation}

\emph{Non-equilibrium magic angles and minimum twist angles}.---We obtain both approximate analytic and exact numerical results of the quasienergy band structures by diagonalizing Eq.~\ref{Eff1} and solving Eq.~\ref{timeE} respectively. We show the optically induced topological flat bands emerge as the driving frequency decreases for certain driving amplitude, see both analytic and numerical results in Fig.~\ref{fig1}(b). The isolated flat bands can also be seen in Fig.~\ref{fig2}(b) and~\ref{fig2}(c) without displacement electric field. As one further increases the driving amplitude for a certain driving frequency or decreases the driving frequency for a certain driving amplitude, the two central bands will touch at the $K^{\prime}$ point. In that case, a topological phase transition happens, and the topological non-trivial bands become trivial bands, see Fig.~\ref{fig2}(d).
 
We adopt three criteria for defining the NEMA: topology, isolation, and bandwidth threshold~\footnote{We set two criteria for the second and fourth NEMA: topology and isolation. This is due to the loss of the maximum bandwidth boundary of the second NEMA. It is because the topological transition from the second to third NEMA happened between the first and second positive bands, which is different from the topological transition at the first NEMA that happened between the two central bands. The isolated region of the fourth NEMA is too narrow to define a maximum bandwidth boundary.}. For the first and third range of NEMA, the range of NEMA is defined by a bandwidth smaller than $13$ meV and $3$ meV, respectively, see Fig.~\ref{fig3}(a) and~\ref{fig3}(b). The minimum bandwidth we estimated is $12.5$ meV for the first range of NEMA. The second and fourth ranges of NEMA are also located at the regions where the bandwidth becomes minimal and can be traced back to MTA, see Fig.~\ref{fig3}(a) and~\ref{fig3}(b). We can use the band gap $\delta_2$ between the first positive central band and the second positive band to distinguish them, as the $\delta_2>0$ the first positive central band becomes isolated, see Fig.~\ref{fig3}(e) and~\ref{fig3}(f). The band gaps can be seen in Fig.~\ref{fig3}(e),~\ref{fig3}(f), and~\ref{fig3}(g) which are positive indicating the flat bands are isolated.

The origin of the NEMA can be traced back to the MTA in the static cases. In the equilibrium situation, they are located at the minimum values of bandwidth which are not isolated (the band gaps $\delta_1 = 0$ and $\delta_2 < 0$), see Fig.~\ref{fig3}(d). One can also see them as the driving amplitude or period is decreased to zero~\footnote{The period $T$ approaches zero or the driving frequency becomes infinity corresponds to the case of fast driving limit. In our system, the Floquet Hamiltonian in the fast driving limit is the same as the static Hamiltonian because the time average of the time-dependent Hamiltonian is the same as the static Hamiltonian in the continuum model of the Dirac system.}, in that case, the system goes back to the limit of equilibrium, see Fig.~\ref{fig3}(a) and~\ref{fig3}(b) which are pointed by down arrows. We estimate the first four MTA as $\theta_1=1.19$, $\theta_2=0.75$, $\theta_3=0.55$, and $\theta_4=0.41$. The same MTA is obtained in AA/AB and AA/ABC stacked TMG (Supplemental Material).

\emph{Displacement electric field}.---We apply the displacement electric field to the system. The electric field can largely flatten the band without causing the bands to lose the topological and isolation properties. The minimum bandwidth at the first range of non-equilibrium magic angles is reduced from $12.5$ meV to 7.8 meV and $6.8$ meV as the electric field $\Delta V$ reaches $200$ meV and $400$ meV separately, see Fig.~\ref{fig4}(a) and~\ref{fig4}(b). The band gaps are positive and can be seen in Fig.~\ref{fig4}(c) and~\ref{fig4}(d) showing the flat band is isolated. We estimate the first range of NEMA from $1.04^\circ$ to $1.45^\circ$ for $\Delta V=200$ meV and from $1.1^\circ$ to $1.68^\circ$ for $\Delta V=400$ meV. The first range of NEMA is enlarged and controlled by the electric field.

An applied electric field has been previously shown to have the ability to engineer the central flat bands without laser in other stacked TMG~\cite{Zhang2024}. The combined effect of laser and displacement electric field in this paper presents the emergence of more flattened bands, see Fig.~\ref{fig3}(h).
\begin{figure}[ht]
   \centering
   \includegraphics[width=3.4in]{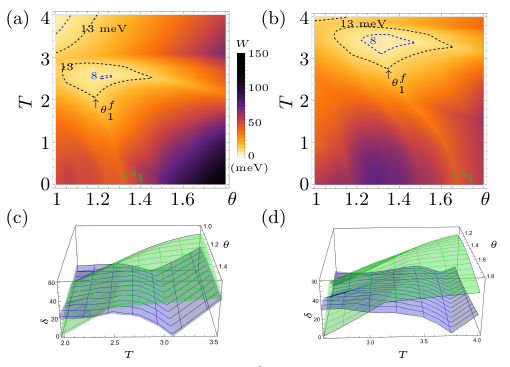} 
   \caption{Exact numerical solutions. (a) and (b) show the quasienergy bandwidth ($W$) of the first positive central band with displacement electric field $\Delta V = 200$ and $400$ meV separately. $\theta^{f}_{1}$ is in the first range of non-equilibrium magic angles (pointed by up arrows) and $\theta_{1}$ is the twisted angle (pointed by down arrows) of the corresponding static origin. In (a) and (b), $A_0=0.45$. (c) and (d) show the quasienergy band gaps $\delta$ between the two central bands (blue) and between the first positive band and the second positive band (green). (c) and (d) are corresponding (a) and (b) separately.}
   \label{fig4}
\end{figure}

\emph{Floquet Chern number}.---
We use the following Hamiltonian to calculate the Floquet Chern number
\begin{equation}\label{Eff2}
\begin{aligned}
 H =&\begin{bmatrix}
H^{(0)}-\Omega & H^{(1)} & 0\\
H^{(-1)} &H^{(0)} &H^{(1)}\\
0 & H^{(-1)} &H^{(0)}+\Omega \\
\end{bmatrix}.
\end{aligned}
\end{equation}
The topological Floquet Chern numbers are numerically calculated~\cite{Fukui2005}. It is defined in the moir\'e Brillouin zone as follows
\begin{equation}
    C=\frac{1}{2 \pi }\int_{\text{mBZ}} F_{m,n}(\bold{k}) d^{2}\bold{k}, 
\end{equation}
where
\begin{equation}
    F_{m,n}(\bold{k})=\partial_{k_m}B_{n}(\bold{k})-\partial_{k_n}B_{m}(\bold{k}),
\end{equation}
and Berry connection is defined with Floquet wave functions as~\cite{Oka2009}
\begin{equation}
    B_{m}(\bold{k})=i\sum_{i}\langle\phi_{\bold{k}}^{(i)}|\partial_{k_m}\ket{\phi_{\bold{k}}^{\left(i\right)}}.
\end{equation}

We obtain the Floquet Chern number $C=\pm 1$, $C=\pm 2$, $C=\pm 1$, and $C=\pm 2$ (per spin per valley) of the two central topological bands for the four ranges of NEMA (from the first to the fourth range of NEMA) accordingly. Since the moir\'e bands are in the very low energy range, the Floquet sidebands do not exist in this range~\cite{Li_2020}. We estimate amplitude of $\phi_{\bold{k}}^{(1)}$ and $\phi_{\bold{k}}^{(-1)}$ are very small ($\approx1\%$) compared to the amplitude of $\phi_{\bold{k}}^{(0)}$. 

\emph{Discussion}.---The bandwidth of the isolated optically induced topological flat bands we found is $\sim 12.5$ meV without the electric field. This is comparable to the bandwidth of TBG, which exhibited correlated insulating and superconducting states below the first flat-band magic angle with a bandwidth of $\sim 11$ meV~\cite{Codecido2019}. It is therefore reasonable to expect similar exotic physics~\cite{Liu2018,Arakawa2021,Zhang2015,Takasan2017,Cayao2021} to emerge from the flat bands we reported in this work. As the displacement electric field increases, the bandwidth can be further reduced, producing Landau level-like flat bands that could potentially host Floquet fractional Chern insulating states ~\cite{Grushin2014}. Therefore, the optically induced topological flat bands make AA-stacked TMG an attractive platform to explore such exotic dynamic effects.

In addition, AA-stacked bilayer graphene can be found naturally. Related experiments have also been conducted with a tiny twist angle of approximately $0.1^{\circ}$ to create a large region of AA-like stacking~\cite{Kim2013}. Although the low-energy band structure in such a system is modified by a tiny twist and an electric field, the Fermi ring formed by the shift of two Dirac cones can still be partially observed, preserving the main structure of AA-stacked bilayer graphene. This opens up opportunities for experiments on AA-stacked TMG. This can further enable the exploration of the optically induced topological flat bands we discovered in AA-stacked TMG. Additionally, recent experiments~\cite{Marco2024,choi2024} observing Floquet states in single-layer graphene could be relevant to potential optical experiments on TBG for investigating Floquet states, paving the way for similar studies on AA-stacked TMG.

\emph{Summary}.---We uncover a series of non-equilibrium magic angles in illuminated AA-stacked TMG with isolated optically induced topological flat bands. In contrast to previously reported optically induced topological flat bands in the literature, the flat bands we found are not inherited from the flat bands in equilibrium, but originate from the bands with minimal bandwidths. A displacement electric field can additionally tune the flatness of the topological bands and therefore enlarges the range of the magic angles. We believe that the electrically tunable non-equilibrium flat bands can stimulate interest in exploring the exotic physics in the AA-stacked TMG.

\begin{acknowledgments}
This work was supported by the U.S. Department of Energy, Office of Science, Basic Energy Sciences under Early Career Award No. DE-SC0019326. 
\end{acknowledgments}

\bibliography{ref}

\vspace{-5mm}
\onecolumngrid
\newpage

\renewcommand{\thefigure}{S\arabic{figure}}
\renewcommand{\theequation}{S\arabic{equation}}
\setcounter{equation}{0}
\setcounter{figure}{0}

\title{
Supplemental Material for ``Tunable Non-Equilibrium Magic and Minimum Twist Angles in AA-Stacked Twisted Multilayer Graphene''
}

\begin{abstract}
In this supplemental material, we present more details on the AA/S structure discussed in the main text. Also, we show the results of AA/AB and AA/ABC stacked TMG structures. It turns out the AA/AB and AA/ABC structures have the same MTA as the AA/S structure in the main text. The NEMA are also similar to the NEMA in AA/S showing both NEMA and MTA are universal in AA-stacked TMG.
\end{abstract}

\date{\today}

\maketitle


\onecolumngrid
\section{AA/S stacked twisted multilayer graphene}
\noindent
\subsection{Mechanism of MTA for the isolated flat bands}
\begin{figure}[H]
   \centering
   \includegraphics[width=5.2in]{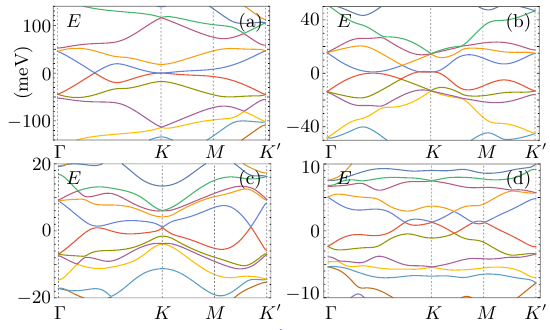} 
   \caption{Exact numerical solutions for AA/S structure with trigonal warping. (a) is at first MTA $\theta_1=1.19^\circ$. (b) is at the second MTA $\theta_2=0.75^\circ$. (c) is at the third MTA $\theta_3=0.55^\circ$. (d) is at the fourth MTA $\theta_4=0.41^\circ$. Note that the particle-hole asymmetry is due to the trigonal warping and lattice relaxation $u$.}
   \label{FigS1}
\end{figure}

The band structure of the first four MTA of the AA/S structure can be seen in Fig.~\ref{FigS1}. The MTA appears when the values of symmetry point $\Gamma$ and $K^{\prime}$ of the two central bands are equal, see Fig.~\ref{FigS1}. This is because the energy at the $\Gamma$ point is mainly from the electrons of the single-layer graphene (S) and the energy at the $K^{\prime}$ point is primarily from the AA bilayer graphene (AA). The two energies are the same as the twist angles reach the MTA. The competition of the two energies provides the condition to form isolated flat bands in non-equilibrium. This mechanism of appearance of MTA is also applicable to other structures of AA stacked TMG, such as AA/AB and AA/ABC, see Fig.~\ref{FigS4} in the following sections. 

To compare the origin mechanism with flat-band magic and Dirac magic, we note that, in AA-stacked TMG, the flat bands can not form by setting the lattice relaxation $u=w_{\text{AA}}/w_{\text{AB}}=0$ which is true at the flat-band magic angles of the twisted bilayer graphene~\cite{Bistritzer_2011,Tarnopolsky_2019} indicating the MTA has a different mechanism forming flat bands. Also, the Dirac magic has the origin by setting interlayer hopping $w_{\text{AB}}=0$~\cite{Li2022} which does not form the MTA indicating the MTA has a different mechanism from Dirac magic.
\subsection{Isolated extremely flat bands}
\begin{figure}[ht]
   \centering
   \includegraphics[width=5.2in]{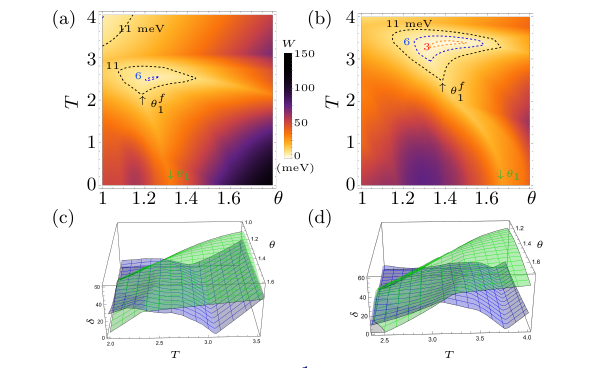} 
   \caption{Exact numerical solutions. (a) and (b) show the quasienergy bandwidth ($W$) of the first positive central band for the path $\Gamma \rightarrow K\rightarrow M \rightarrow K^{\prime}$ with displacement electric field $\Delta V = 200$ and $400$ meV separately. $\theta^{f}_{1}$ is in the first range of non-equilibrium magic angles (pointed by up arrows) and $\theta_{1}$ is the twisted angle (pointed by down arrows) of the corresponding static origin. In (a) and (b), $A_0=0.45$. The unit of $A_0$ is $\hbar/ea$. the driving frequency $\Omega=2\pi/T$ and the unit of $\Omega$ is $\hbar v_F / a = 2.4$ eV. (c) and (d) show the quasienergy band gaps $\delta$ between the two central bands (blue) and between the first positive band and the second positive band (green). (c) and (d) are corresponding (a) and (b) separately.}
   \label{figS2}
\end{figure}

\begin{figure}[ht]
   \centering
   \includegraphics[width=5.2in]{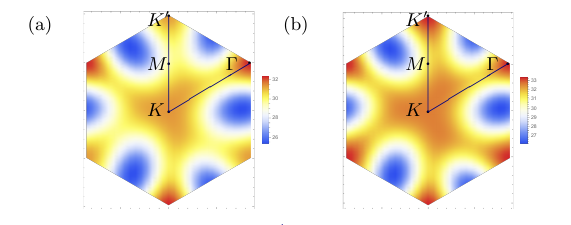} 
   \caption{Exact numerical solutions. Bandwidth (meV) in the entire moir\'e Brillouin zone with displacement electric field $\Delta V = 400$ meV. (a) is at the $\theta^{f}_{1}=1.35^\circ$ and (b) is at the $\theta^{f}_{1}=1.4^\circ$. At the $\theta^{f}_{1}=1.4^\circ$, for the path $\Gamma \rightarrow K\rightarrow M \rightarrow K^{\prime}$, the flat band is extremely flatten and $W\approx 2$ meV. We set $A_0=0.45$ and $T=3.38$.}
   \label{figS3}
\end{figure}

\begin{figure}[H]
   \centering
   \includegraphics[width=5.2in]{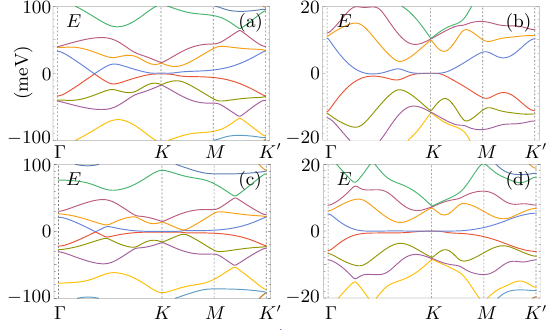} 
   \caption{Exact numerical solutions for AA/AB and AA/ABC structure without trigonal warping. (a) and (b) show the AA/AB band structures at first MTA $\theta_1=1.19^\circ$ and second MTA $\theta_2=0.75^\circ$ separately. (c) and (d) show the AA/ABC band structures at first MTA $\theta_1=1.19^\circ$ and second MTA $\theta_2=0.75^\circ$ separately. Note that the particle-hole asymmetry is due to the lattice relaxation $u$.}
   \label{FigS4}
\end{figure}

For certain path $\Gamma \rightarrow K\rightarrow M \rightarrow K^{\prime}$, the flat band becomes extremely flat under the displacement electric field. When $\Delta V=200$ meV, the bandwidth becomes $2$ meV, see Fig.~\ref{figS2}, while for the entire moir\'e Brillouin zone, it is $6.8$ meV as shown in the main text. We also show the bandwidth in the entire moir\'e Brillouin zone here, see Fig.~\ref{figS3}. Note that the values of MTA do not change for this chosen path compared to the entire moir\'e Brillouin zone.

\section{AA/AB stacked twisted multilayer graphene}
\noindent
The Hamiltonian of AA/AB can be written as ~\cite{Bistritzer_2011,Li2022}
\begin{equation}
H_{\text{AA/D}}=
\begin{bmatrix}
 h_{\text{AA},\theta/2} & T \\
T^\dagger & h_{\text{D},-\theta/2} \\
\end{bmatrix},
\end{equation}
where
\begin{equation}
h_{\text{AA},\theta/2}=
\begin{bmatrix}
 h_{\theta/2} & T_{\text{AA}} \\
T_{\text{AA}}^\dagger & h_{\theta/2} \\
\end{bmatrix}
\end{equation}
and
\begin{equation}
h_{\text{D},-\theta/2}=
\begin{bmatrix}
 h_{-\theta/2} & T_{\text{AB}} \\
T_{\text{AB}}^\dagger & h_{-\theta/2} \\
\end{bmatrix}.
\end{equation}
The interlayer tunneling matrix for AA-stacked bilayer graphene is~\cite{Jung2014} 
\begin{equation}\label{eq:TAA}
T_{\text{AA}}=
\begin{bmatrix}
\gamma^1_{\text{AA}} & \gamma^4_{\text{AA}}\gvex k \\
\gamma^{4}_{\text{AA}}\gvex k ^{\star} & \gamma^1_{\text{AA}} \\
\end{bmatrix},
\end{equation}
and for AB-stacked bilayer graphene is 
\begin{equation}\label{eq:TAB}
T_{\text{AB}}=
\begin{bmatrix}
\gamma^1_{\text{AB}} \gvex k & \gamma^3_{\text{AB}} \gvex k^{\star}  \\
\gamma^4_{\text{AB}} & \gamma^1_{\text{AB}} \gvex k \\
\end{bmatrix},
\end{equation}
where $\gvex k=k_{x}+i k_{y}$ and these $\gvex k$ dependent tunneling matrix elements represent trigonal warping and particle-hole asymmetry. 
We use the parameters as following, $a= 2.4$~\AA, $\hbar v_F / a = 2.425$~eV, and $w_{\text{AB}} = 110$~meV. For AA-stacked bilayer graphene, $\gamma^1_{\text{AA}}=217$ meV and $\gamma^4_{\text{AA}}/a=-20$ meV. For AB-stacked bilayer graphene, $\gamma^1_{\text{AB}}/a=-138$ meV, $\gamma^3_{\text{AB}}/a=-283$ meV, and $\gamma^4_{\text{AB}}=-361$ meV.

\begin{figure}[H]
   \centering
   \includegraphics[width=5.2in]{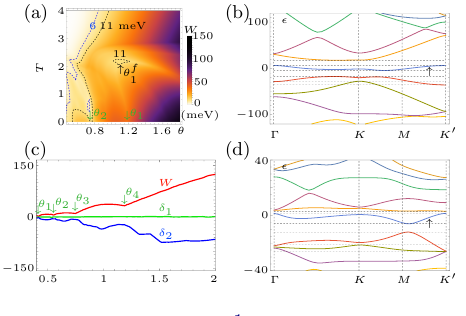} 
   \caption{Exact numerical solutions for AA/AB structure for the path $\Gamma \rightarrow K\rightarrow M \rightarrow K^{\prime}$. (a) quasienergy bandwidth depends on the twisted angle and driving period with trigonal warping. (b) and (d) show the quasienergy band structures in the first and second range of non-equilibrium magic angles with trigonal warping and the topological flat bands are pointed by arrows. (c) shows the static bandwidth and gaps for the first positive central band without trigonal warping in the AB bilayer. In (b), $A_0=0.45$, $T=2.1$, and $\theta^{f}_{1}=1.1^\circ$.  In (d), $A_0=0.45$, $T=1.29$, and $\theta^{f}_{2}=0.68^\circ$. The unit of $A_0$ is $\hbar/ea$. the driving frequency $\Omega=2\pi/T$ and the unit of $\Omega$ is $\hbar v_F / a = 2.4$ eV. }
   \label{FigS5}
\end{figure}

Same as the case of AA/S, we introduce the periodic laser potential in BM model~\cite{Bistritzer_2011} then $h_{\theta/2}(t) = v_F[-i\hbar \gvex\nabla - e \vex A(t)]\cdot\gvex\sigma_{\theta/2}$. We show that the AA/AB structure has the same MTA as the AA/S structure without trigonal warping in the AB bilayer, see Fig.~\ref{FigS5}(c). The first range of NEMA is shifted to smaller magic angles compared to the case of AA/S. We note that the lower range of NEMA is affected by trigonal warping of the AB bilayer and only the first and second ranges of NEMA are kept, see Fig.~\ref{FigS5}(a). The optically induced isolated topological flat bands can be seen in Fig.~\ref{FigS5}(b) and (d). The Floquet Chern numbers are $C=\pm1$ and $C=\pm 3$ for the two central bands in the first and second range of NEMA respectively. Note that only the first positive flat band is isolated in the AA/AB structure, see Fig.~\ref{FigS5}(b) and (d) (pointed by the black arrows).

\section{AA/ABC stacked twisted multilayer graphene}
\noindent
The Hamiltonian of AA/ABC can be written as ~\cite{Bistritzer_2011,Li2022}
\begin{equation}
H_{\text{AA/D}}=
\begin{bmatrix}
 h_{\text{AA},\theta/2} & T \\
T^\dagger & h_{\text{D},-\theta/2} \\
\end{bmatrix},
\end{equation}
where
\begin{equation}
h_{\text{AA},\theta/2}=
\begin{bmatrix}
 h_{\theta/2} & T_{\text{AA}} \\
T_{\text{AA}}^\dagger & h_{\theta/2} \\
\end{bmatrix}
\end{equation}
and
\begin{equation}
h_{\text{D},-\theta/2}=
\begin{bmatrix}
 h_{-\theta/2} & T_{\text{AB}} & T_{\text{AC}}\\
T_{\text{AB}}^\dagger & h_{-\theta/2} & T_{\text{AB}}\\
T_{\text{AC}}^\dagger & T_{\text{AB}}^\dagger & h_{-\theta/2}\\
\end{bmatrix}.
\end{equation}
The interlayer tunneling matrix  without trigonal warping for AA-stacked bilayer graphene is~\cite{Jung2014} 
\begin{equation}\label{eq:TAA}
T_{\text{AA}}=
\begin{bmatrix}
\gamma^1_{\text{AA}} & 0 \\
0 & \gamma^1_{\text{AA}} \\
\end{bmatrix},
\end{equation}
and for AB-stacked bilayer graphene without trigonal warping  is 
\begin{equation}\label{eq:TAB}
T_{\text{AB}}=
\begin{bmatrix}
0 & 0 \\
\gamma^4_{\text{AB}} & 0\\
\end{bmatrix},
\end{equation}
and $T_{AC}=\gamma_{AC}\cdot\sigma^{+}$, where $\sigma^{+}$ is the raising operator. We set $\gamma_{AC}=0$, since it is too small compared to $\gamma^1_{\text{AA}}$ and $\gamma^4_{\text{AB}}$. The parameters used are the same as those in the AA/AB.

\begin{figure}[H]
   \centering
   \includegraphics[width=5.2in]{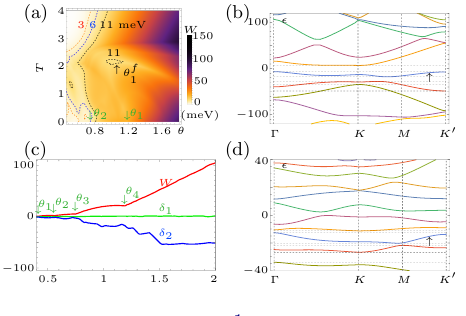} 
   \caption{Exact numerical solutions for AA/ABC structure for the path $\Gamma \rightarrow K\rightarrow M \rightarrow K^{\prime}$. (a) shows quasienergy bandwidth depends on the twisted angle and driving period without trigonal warping. (b) and (d) show the quasienergy band structures in the first and second range of non-equilibrium magic angles without trigonal warping and the topological flat bands are pointed by arrows. (c) shows the static bandwidth and gaps for the first positive central band without trigonal warping in the AB bilayer. In (b), $A_0=0.45$, $T=2.14$, and $\theta^{f}_{1}=1.03^\circ$.  In (d), $A_0=0.45$, $T=2.1$, and $\theta^{f}_{2}=0.5^\circ$. The unit of $A_0$ is $\hbar/ea$. the driving frequency $\Omega=2\pi/T$ and the unit of $\Omega$ is $\hbar v_F / a = 2.4$ eV. }
   \label{FigS6}
\end{figure}

Same as the case of AA/S and AA/AB, we introduce the periodic laser potential in BM model~\cite{Bistritzer_2011} then $h_{\theta/2}(t) = v_F[-i\hbar \gvex\nabla - e \vex A(t)]\cdot\gvex\sigma_{\theta/2}$. We show that the AA/ABC structure has the same MTA as the AA/S and AA/AB structures without trigonal warping in the AA and AB bilayer, see Fig.~\ref{FigS6}(c). The first range of NEMA is shifted to smaller magic angles compared to the case of AA/S and AA/AB. The optically induced isolated topological flat bands can be seen in Fig.~\ref{FigS6}(b) and (d) (pointed by the black arrows). The Floquet Chern numbers are $C=\pm1$ and $C=\pm 2$ for the two central bands in the first and second range of NEMA respectively. 

\end{document}